\documentclass[cameraready]{Interspeech}
\usepackage{amsmath,graphicx,hyperref}
\usepackage{amssymb}
\usepackage{xcolor}
\usepackage{subcaption}
\usepackage{url}
\usepackage[dvipsnames]{xcolor}
\usepackage{hyperref}
\usepackage{booktabs}
\usepackage{multirow}
\usepackage{float}
\usepackage{enumitem}

\usepackage{siunitx} % For S column type, ensuring perfect alignment

\definecolor{improvementcolor}{HTML}{6495EC} % A nice blue
\definecolor{degradationcolor}{HTML}{B12122} % A nice red

\newcommand{\improvement}[1]{\textcolor{improvementcolor}{\num{#1}}}
\newcommand{\degradation}[1]{\textcolor{degradationcolor}{\num{#1}}}

\title{Data-Efficient ASR Personalization for Non-Normative Speech Using an Uncertainty-Based Phoneme Difficulty Score for Guided Sampling\thanks{Under review at Interspeech 2026.}}

\author[affiliation={1,2}, orcid=0009-0005-6429-2019]{Niclas}{Pokel}
\author[affiliation={1}, orcid=0009-0001-1631-8238]{Pehuén}{Moure}
\author[affiliation={1}, orcid=0000-0003-2856-3262, correspondingauthor]{Roman}{Böhringer}
\author[affiliation={3}, orcid=0009-0000-0876-621X, correspondingauthor]{Yingqiang}{Gao}

\address{
$^1$ Institute of Neuroinformatics, University of Zurich and ETH Zurich, Switzerland \\
$^2$ School of Computation, Information and Technology, Technical University of Munich, Germany \\
$^3$ Department of Computational Linguistics, University of Zurich, Switzerland
}

\email{ \{npokel, pehuen, roman\}@ini.ethz.ch, yingqiang.gao@cl.uzh.ch}


\usepackage{comment}

\begin{document}

\maketitle

% the abstract here must exactly match the abstract entered into the paper submission system
\begin{abstract}
    % 1000 characters. ASCII characters only. No citations.
ASR systems struggle with non-normative speech due to high acoustic variability and data scarcity. We propose a data-efficient method using phoneme-level uncertainty to guide fine-tuning for personalization. Instead of computationally expensive ensembles, we leverage Variational Low-Rank Adaptation (VI LoRA) to estimate epistemic uncertainty in foundation models. These estimates form a composite Phoneme Difficulty Score (PhDScore) that drives a targeted oversampling strategy. Evaluated on English and German datasets, including a longitudinal analysis against two clinical reports taken one year apart, we demonstrate that: (1) VI LoRA-based uncertainty aligns better with expert clinical assessments than standard entropy; (2) PhDScore captures stable, persistent articulatory difficulties; and (3) uncertainty-guided sampling significantly improves ASR accuracy for impaired speech.
\end{abstract}

\section{Introduction}
\label{sec:intro}

Despite significant progress in automatic speech recognition (ASR), state-of-the-art models like Whisper \cite{radford2023robust} still fail when processing non-normative utterances from individuals with speech impairments \cite{Rowe2022, hustad2019differentiating}. The challenge is especially pronounced in children, whose speech patterns evolve dynamically \cite{long2022speech}, and for languages like German that lack sufficient non-normative training data \cite{Guldimann2024}. Fine-tuning pre-trained models is a common personalization strategy \cite{Leivaditi2024, Zheng2024, shor19_interspeech}, but with limited per-individual data it is prone to overfitting. Other data-efficient techniques such as data augmentation \cite{leung2024training, soleymanpour2024accurate} or parameter-efficient fine-tuning \cite{hu2022lora, qi23b_interspeech} typically treat all training samples equally, missing an opportunity to focus on problematic speech patterns.
The field of confidence estimation in ASR has established methods to gauge prediction reliability \cite{Li2021Confidence, Woodward2020Confidence, Kumar2020Utterance}, typically used in post-processing for error detection \cite{Kuhn_2025, Shu2024Error}. We instead propose an uncertainty-based score that steers the model's focus during training itself. While softmax-based uncertainty is efficient \cite{Park2020ImprovedNoisyStudent}, it is often unreliable \cite{Hendrycks2017Baseline, Li2021Confidence}. Bayesian Neural Networks offer a robust alternative in low-data settings \cite{moure2024regularized}, commonly via Monte Carlo Dropout (MCD) \cite{NIPS2017_2650d608, gal2016dropout}, but MCD is expensive for large Transformers \cite{xiao2019wat} and, as we show, simple entropy metrics fail to distinguish acoustic noise from specific articulatory difficulties. We therefore utilize Variational Low-Rank Adaptation (VI~LoRA) \cite{pokel2025vilora} to efficiently estimate epistemic uncertainty and compute a novel Phoneme Difficulty Score (PhDScore) as a fine-grained proxy for a speaker's impairments.
Recent work has demonstrated the value of phoneme-level analysis for dysarthric speech through contrastive learning with phonetic difficulty curricula \cite{lee2025dypcldynamicphonemelevelcontrastive}, model-level phoneme confusion \cite{YUAN2026132684} and pronunciation features for severity classification \cite{Yeo21Interspeech}, but these focus on representations or assessment rather than guiding training data distribution. General strategies for prioritizing samples exist, curriculum learning \cite{Curriculum, hsieh24_interspeech}, learned reweighting \cite{ren2018learning}, focal loss \cite{Lin_2017_ICCV}, but assume sufficient training data or class imbalance and lack a domain-specific signal for impaired speech difficulty. Oversampling has been applied to dysarthric ASR for class balancing \cite{ASRreweight}, uncertainty sampling explored for active learning \cite{deng21_interspeech}, and hybrid selection combining diversity with transcription confidence \cite{hiruta25_interspeech}. Entropy-based weighting also has a long history in ASR \cite{misra2003}. In contrast, our method derives a clinically grounded difficulty signal from phoneme-level uncertainty to strategically re-weight a small, pre-existing dataset. Our contributions are:

\begin{enumerate}[noitemsep, topsep=0pt, partopsep=0pt, parsep=0pt]
    \item \textbf{A composite uncertainty-based metric for estimating phoneme difficulty.} We formalize a score combining multiple uncertainty metrics to identify challenging phonemes more robustly than entropy alone.
    \item \textbf{Efficient uncertainty-guided oversampling.} We introduce a BNN–based training strategy that targets the hardest acoustic patterns and yields direct epistemic uncertainty estimates via Bayesian adapters, without masking representations.
    \item \textbf{Longitudinal clinical validation.} We demonstrate our method's effectiveness on English and German datasets and show that the PhDScore strongly correlates with two clinical logopedic reports taken one year apart.
\end{enumerate}

\section{Methodology}
\label{sec:method}

Our framework creates a quantitative proxy for clinical speech difficulty to guide data-efficient personalization of ASR models. This involves three steps: (1) estimating model uncertainty via stochastic forward passes, (2) computing the PhDScore, and (3) oversampling difficult utterances for fine-tuning. We investigate two methods for uncertainty estimation: Monte Carlo Dropout (MCD) and Variational Low-Rank Adaptation (VI LoRA).

\subsection{Uncertainty Estimation: Monte Carlo Dropout (MCD)}
Following \cite{gal2016dropout}, we treat dropout at inference time as a Bayesian approximation. We inject dropout layers with rate $p_{\text{drop}}=1\%$ after the first two linear layers of each Transformer block in the Whisper backbone. Uncertainty results were stable for $0.2\% <p_{\text{drop}}<3\%$, while  $p_{\text{drop}}>3\%$ led to model collapse for transcription due to cascading effects of dropout in large models. To estimate uncertainty for an input $x$, we perform $M=20$ stochastic forward passes with dropout. This yields an ensemble of predictions generated by implicit sub-models. Marginal changes in uncertainty estimates were observed for $M>20$, while $M<10$ produced unstable rankings.

\subsection{Uncertainty Estimation: Variational Low-Rank Adaptation (VI LoRA)}
VI LoRA \cite{pokel2025vilora} extends standard LoRA \cite{hu2022lora} by modeling the adapter matrices $A$ and $B$ as variational distributions $q_\phi(A)$ and $q_\phi(B)$ rather than fixed weights. We use a mean-field approximation with diagonal Gaussians:
\begin{align}
    q_{\phi}(A) &= \prod_{i,j} \mathcal{N}(A_{ij}| \mu_{A_{ij}}, \sigma^2_{A_{ij}}), \nonumber \\
    q_{\phi}(B) &= \prod_{k,l} \mathcal{N}(B_{kl}|\mu_{B_{kl}}, \sigma^2_{B_{kl}}).
\end{align}
Parameters $\phi$ are learned by minimizing the negative Evidence Lower Bound (ELBO), using a bimodal prior $p(A,B)$ derived from pre-trained weights \cite{pokel2025vilora}. During inference, stochasticity is induced by sampling adapter weights $A^{(m)} \sim q_{\phi}(A)$ and $B^{(m)} \sim q_{\phi}(B)$ for $m$ being each of the $M$ passes. This restricts stochasticity to the parameter-efficient adapters, leaving the massive backbone deterministic.

\subsubsection{Predictive Uncertainty Calculation}
For both methods, we quantify predictive uncertainty as the entropy over the vocabulary $\mathcal{V}$ of the aggregated predictive distribution. Let $\hat{p}(j)$ be the averaged probability of token $j$ across $M$ passes:
\begin{align}
    & \hat{p}(j) = \frac{1}{M} \sum_{m=1}^M p(y=j \, | \, x, \theta^{(m)}) \nonumber, \\
    & H(y|x) \approx - \sum_{j \in \mathcal{V}} \hat{p}(j) \log_2 \hat{p}(j), 
    \label{eq:entropy}
\end{align}
where $\theta^{(m)}$ represents either the dropout-masked parameters or the sampled VI LoRA adapters.

\subsection{Composite Phoneme Difficulty Score (PhDScore)}
\label{subsec:phdscore}

We found that entropy alone is insufficient to capture clinical difficulty (see Sec.~\ref{sec:results}). Thus, we formulate a composite \textbf{PhDScore} for each phoneme type $p$. We aggregate three normalized metrics over all instances of phoneme $p$ (denoted as set $I_p$) in the user's data:

\begin{enumerate}[leftmargin=*]
    \item \textbf{Phoneme Error Rate ($E_p$):} The ratio of incorrect majority-vote ($\text{maj}$) predictions to total occurrences.
    \item \textbf{Mean Prediction Entropy ($H_p$):} The average predictive entropy (Eq.~\ref{eq:entropy}) across instances (min-max normal. to $[0,1]$).
    \item \textbf{Ground Truth Agreement ($A_p$):} The frequency with which stochastic samples match the ground truth $\text{gt}_i$.
\end{enumerate}

\noindent These metrics are calculated as follows:
\begin{align}
    E_p &= \frac{\text{Count-Error}_{\text{maj}}(p)}{|I_p|}, \quad 
    H_p = \frac{1}{|I_p|} \sum_{i \in I_p} H(y_i|x_i), \nonumber \\
    A_p &= \frac{1}{|I_p|} \sum_{i \in I_p} \left( \frac{1}{M} \sum_{k=1}^{M} \mathbb{I}(\text{pred}_{i,k} = \text{gt}_i) \right).
\end{align}
The final PhDScore is a weighted sum, where $A_p$ is inverted as high agreement implies low difficulty:
\begin{equation}
\text{PhDScore}_p = w_e E_p^{\text{norm}} + w_h H_p^{\text{norm}} + w_a (1 - A_p^{\text{norm}}).
\end{equation}
Weights were determined via a grid search on a validation subset. While results were found to be robust to minor variations in weighting (including equal weighting), we utilized the configuration ($w_e=0.4, w_h=0.2, w_a=0.4$) to prioritize discrete error and agreement signals over the noisier entropy metric.

\subsection{Uncertainty-Guided Oversampling}
To guide fine-tuning, we aggregate phoneme-level PhDScores into an utterance-level weight by averaging the scores of all constituent phonemes. The utterance-level scores are min-max normalized across the training set to a sampling probability range of $[1.0, 5.0]$. Intuitively, we compute the PhDScore using the pre-trained (zero-shot) model. As the model fine-tunes, its epistemic uncertainty regarding the specific speaker diminishes, making the signal less discriminative for further training.

\section{Experimental Setup}
\label{sec:exp}

We evaluate on UA-Speech \cite{kim2008dysarthric} (English, 16 speakers with varying dysarthria) and BF-Sprache \cite{pokel25_diss} (German, 505 isolated words from a child with Apert syndrome). Semantic re-chaining \cite{pokel25_diss} bridges the gap between these isolated-word datasets and foundation models optimized for continuous speech by concatenating recordings into semantically coherent sentences. To assess generalization and potential forgetting, we also evaluate on Mozilla Common Voice \cite{ardila-etal-2020-common}. 

All experiments use a 70/10/20 train/val/test split with cross-validation and seed variation for confidence intervals. We compare three adaptation strategies: full fine-tuning (Full~FT), LoRA \cite{hu2022lora}, and VI~LoRA \cite{pokel2025vilora}. We use Adam \cite{kingma2017adammethodstochasticoptimization} with default parameters, effective batch size 32, a 10\% relative KL-divergence weight for VI~LoRA, and learning rates of 5e-6 (FT) and 1e-4 (VI~LoRA and LoRA), with early stopping on validation non-normative CER. Experiments ran on dual AMD EPYC 7742 CPUs (128 cores, 256\,GiB RAM) with up to four NVIDIA RTX 3090 GPUs (24\,GiB each) in a shared environment, distributed training used DeepSpeed, with average training times around one hour on 4~GPUs.

\section{Results and Analysis}
\label{sec:results}

\subsection{The Personalization-Generalization Trade-off}
\begin{figure}[t]
    \centering
    \includegraphics[width=\columnwidth]{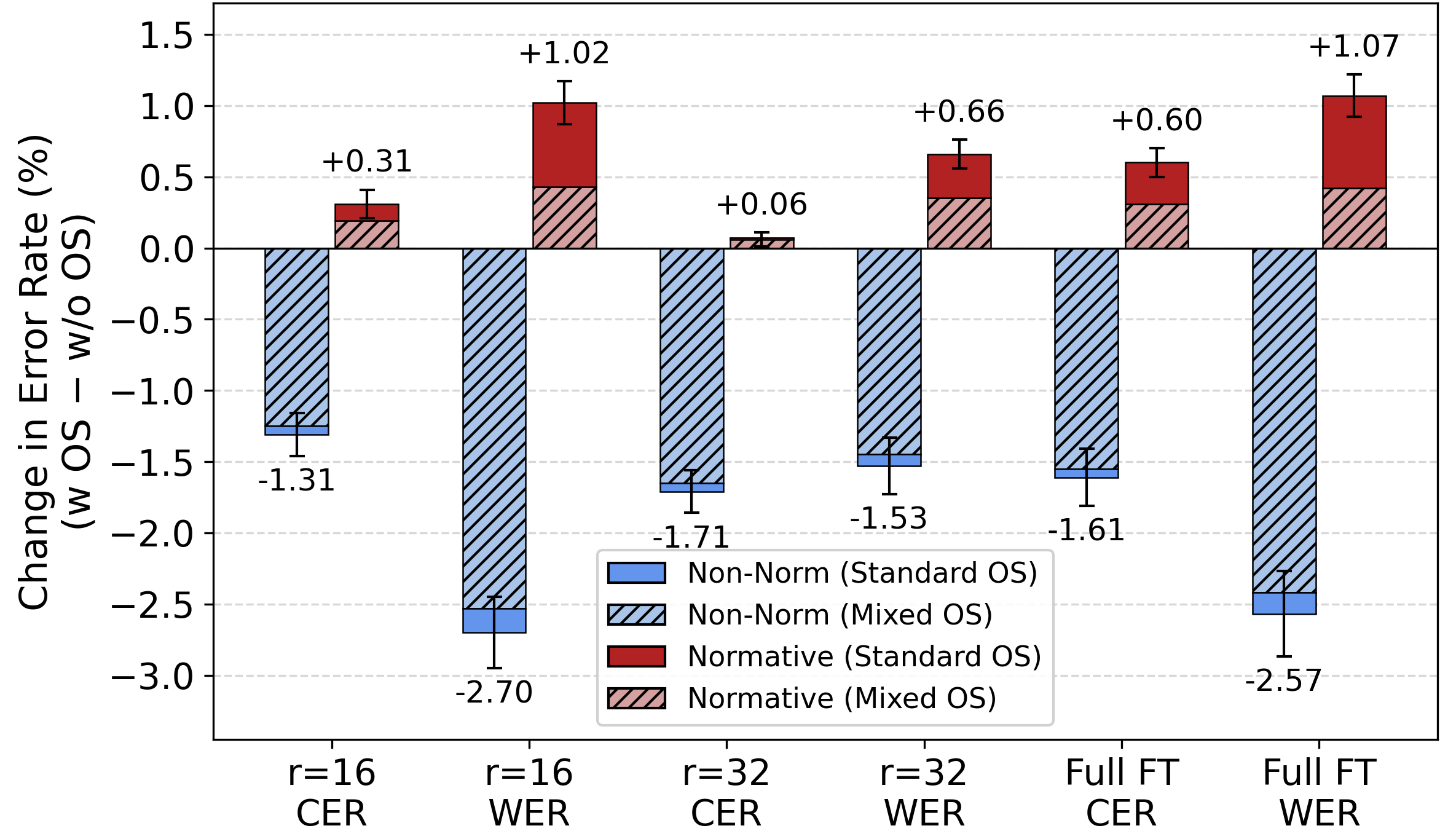}
    \caption{We compare LoRA and full fine-tuning (FT) with (w) against without (w/o) oversampling (OS). Negative values (\textcolor{improvementcolor}{blue}) indicate an improvement on BF-Sprache, while positive values (\textcolor{degradationcolor}{red}) show forgetting of normative speech.}
    \label{fig:oversampling_tradeoff_larger}
\end{figure}

While our method consistently improves performance on the target non-normative speech, it is crucial to quantify the effect this specialization has on the model's ability to transcribe general, normative speech. Figure \ref{fig:oversampling_tradeoff_larger} visualizes this trade-off by plotting the percentage point change in error rates when uncertainty-guided oversampling is applied, compared to a standard training baseline.

\noindent \textbf{Results on BF-Sprache.}
The results reveal a clear personalization-generalization trade-off. For non-normative speech, all blue bars show a negative change, confirming that oversampling effectively reduces error rates by up to 2.70 percentage points in WER (LoRA, $r=16$). However, this comes at a cost: for normative speech, all red bars are positive, indicating a degree of catastrophic forgetting. The LoRA rank allows for tuning this trade-off, with the $r=32$ configuration showing reduced normative degradation while still providing strong personalization. To mitigate this forgetting, we explore a mixed variant that interleaves normative samples into the oversampled training set. As shown by the hatched segments in Figure~\ref{fig:oversampling_tradeoff_larger}, this substantially reduces normative degradation while retaining the majority of the personalization gain, offering a practical path toward deployment without sacrificing generalization.
As observed on BF-Sprache, the mixed oversampling variant consistently reduced normative degradation across intelligibility levels while preserving the personalization benefit.

\noindent \textbf{Results on UA-Speech.}
The results for the entire UA-Speech speaker base, presented in Table \ref{tab:ua_speech_side_by_side_final}, generalize the findings from the BF-Sprache dataset. A clear trend emerges: the efficacy of oversampling appears to be inversely correlated with speaker intelligibility. We note that while standard LoRA frequently yields larger relative gains, VI LoRA consistently achieves lower baseline error rates, leaving less capacity for further improvement via oversampling. This suggests that oversampling is most beneficial for individuals with more severe impairments, likely due to stronger error signals from acoustically challenging phonemes or distinct pathological patterns. Distinguishing between these hypotheses would require expert annotation beyond the scope of this work. Oversampling also led to a slight degradation on normative speech. Notably, this trend was reversed for the lowest-intelligibility cohort, where oversampling improved performance. Although current data cannot fully explain this anomaly, it suggests that this group’s highly atypical yet consistent speaking patterns are readily learned, and that oversampling may function as noise injection that pushes the model to rely more on linguistic priors.
    
\begin{table}[t]
\caption{Impact of uncertainty-guided oversampling on model performance, showing the percentage point (\%) change in error rates. Each method is compared against its own baseline performance \emph{without} oversampling. Negative values (\textcolor{improvementcolor}{blue}) indicate improved performance on UA Speech. Positive values (\textcolor{degradationcolor}{red}) indicate performance forgetting on Normative speech.}
\label{tab:ua_speech_side_by_side_final}
\centering
\sisetup{
  table-align-text-post=false, % Align numbers, not the symbols
  table-number-alignment=center,
  separate-uncertainty=true,
  tight-spacing=true
}
\large
\resizebox{\columnwidth}{!}{
\begin{tabular}{
  l
  l
  S[table-format=-2.2(2)]
  S[table-format=-2.2(2)]
  S[table-format=1.2(2)]
  S[table-format=1.2(2)]
}
\toprule
\multirow{2.5}{*}{\textbf{Intelligibility}} & \multirow{2.5}{*}{\textbf{Setup}} & \multicolumn{2}{c}{\textbf{Non-Normative}} & \multicolumn{2}{c}{\textbf{Normative}} \\
\cmidrule(lr){3-4} \cmidrule(lr){5-6}
& & {$\Delta$CER (\%)} & {$\Delta$WER (\%)} & {$\Delta$CER (\%)} & {$\Delta$WER (\%)} \\
\midrule

% --- High Intelligibility ---
\multirow{3}{*}{High}
& Full FT & \improvement{-0.67} & \improvement{-1.85} & \degradation{+0.60} & \degradation{+2.04} \\
& LoRA      & \improvement{-0.24} & \improvement{-1.73} & \degradation{+3.41} & \degradation{+4.72} \\
& VI LoRA      & \improvement{-1.01} & \improvement{-1.92} & \degradation{+0.17} & \degradation{+1.86} \\
\midrule

% --- Medium Intelligibility ---
\multirow{3}{*}{Medium}
& Full FT & \improvement{-1.40} & \improvement{-2.68} & \degradation{+0.89} & \degradation{+2.40} \\
& LoRA      & \improvement{-2.75} & \improvement{-5.14} & \degradation{+2.51} & \degradation{+4.29} \\
& VI LoRA      & \improvement{-2.71} & \improvement{-4.97} & \degradation{+0.77} & \degradation{+1.46} \\
\midrule

% --- Low Intelligibility ---
\multirow{3}{*}{Low}
& Full FT & \improvement{-2.83} & \improvement{-5.01} & \degradation{+3.08} & \degradation{+4.78} \\
& LoRA      & \improvement{-8.12} & \improvement{-8.35} & \degradation{+0.69} & \degradation{+1.52} \\
& VI LoRA      & \improvement{-6.31} & \improvement{-6.01} & \degradation{+1.72} & \degradation{+3.40} \\
\midrule

% --- Very Low Intelligibility ---
\multirow{3}{*}{Very Low}
& Full FT & \improvement{-4.83} & \improvement{-7.11} & \degradation{+4.46} & \degradation{+6.23} \\
& LoRA      & \improvement{-14.97} & \improvement{-15.12} & \improvement{-1.22} & \improvement{-4.01}\\
& VI LoRA      & \improvement{-11.57} & \improvement{-13.22} & \degradation{+0.79} & \degradation{+2.05} \\
\midrule

% --- Overall Average ---
\multirow{3}{*}{\textbf{Overall}}
& Full FT & \improvement{-2.43} & \improvement{-3.16} & \degradation{+2.11} & \degradation{+3.83} \\
& LoRA      & \improvement{-6.43} & \improvement{-8.70} & \degradation{+1.16} & \degradation{+2.63} \\
& VI LoRA & \improvement{-5.40} & \improvement{-6.53} & \degradation{+0.86} & \degradation{+2.19} \\
\bottomrule
\end{tabular}
}
\end{table}

% \begin{table}[htb]
% \caption{Impact of the Uncertainty Source on Oversampling Performance. Using guidance from the pre-trained model is essential for effective personalization.}
% \label{tab:uncertainty_source_impact}
% \centering
% \begin{tabular}{@{}llcc@{}}
% \hline
% \textbf{Dataset} & \textbf{Model for MCD} & \textbf{CER} & \textbf{WER} \\
% \hline
% \multirow{3}{*}{UA-Speech F02} & Standard Fine-tuning & 9.54 & 11.80 \\
% & Pre-trained Model & \textbf{2.81} & \textbf{5.98} \\
%     & Fine-tuned Model & 9.48 & 11.83 \\
% \hline
% \multirow{3}{*}{BF-Sprache} & Standard Fine-tuning & 22.60 & 46.43 \\
% & Pre-trained Model & \textbf{20.97} & \textbf{43.42} \\
% & Fine-tuned Model & 23.11 & 46.27 \\
% \hline
% \end{tabular}

% \end{table}

\begin{table}[t]
\caption{Impact of the uncertainty signal source (VI LoRA) on oversampling effectiveness (Full FT). Values show $\Delta$ error rate vs. standard fine-tuning. Only the PhDScore from the pre-trained model yields consistent improvements.}
\label{tab:ablation_sampling}
\centering
\footnotesize
\setlength{\tabcolsep}{4pt}
\begin{tabular}{llcccc}
\toprule
\multirow{2.5}{*}{\textbf{Signal Source}} & \multirow{2.5}{*}{\textbf{Metric}} & \multicolumn{2}{c}{\textbf{BF-Sprache}} & \multicolumn{2}{c}{\textbf{UA-Speech}} \\
\cmidrule(lr){3-4} \cmidrule(lr){5-6}
&  & $\Delta$CER & $\Delta$WER & $\Delta$CER & $\Delta$WER \\
\midrule
\multirow{2}{*}{Pre-trained} & Entropy   & \degradation{+0.25} & \improvement{-1.33} & \improvement{-0.71} & \degradation{+0.11} \\
                              & PhDScore & \textbf{\improvement{-1.61}} & \textbf{\improvement{-2.57}} & \textbf{\improvement{-2.43}} & \textbf{\improvement{-3.16}} \\
\midrule
\multirow{2}{*}{Fine-tuned}  & Entropy   & \degradation{+0.55} & \improvement{-0.53} & \improvement{-0.58} & \degradation{+1.21} \\
                              & PhDScore  & \degradation{+0.48} & \improvement{-0.78} & \degradation{+0.63} & \degradation{+0.98} \\
\bottomrule
\end{tabular}
\end{table}

\noindent \textbf{Ablation of Different Signals.}
Table \ref{tab:ablation_sampling} reveals two critical insights regarding the uncertainty signal, consistent across both datasets. First, the composite score is essential. While the PhDScore derived from the pre-trained model yields substantial error reductions on both BF-Sprache ($\Delta$CER $-1.61$) and UA-Speech ($\Delta$CER $-2.43$), raw entropy produces inconsistent results, sometimes even degrading performance. This suggests that entropy often captures unlearnable acoustic noise (aleatoric uncertainty), whereas the PhDScore isolates the epistemic difficulty that can be resolved through targeted training. Second, the signal must come from the pre-trained model. When using uncertainty from an already fine-tuned model (bottom rows), the oversampling provides no consistent benefit across either dataset. This aligns with our longitudinal analysis (Fig.~\ref{fig:clinical_validation_grid}): the fine-tuned model has already adapted to the speaker's patterns, ``resolving'' its epistemic uncertainty and leaving a non-discriminative signal. Having established that uncertainty-guided oversampling consistently improves performance across speakers, intelligibility levels, and languages, we now investigate whether the underlying signal captures clinical difficulty.

\subsection{Validation Against Longitudinal Clinical Assessments}
\label{sec:clinical}
\begin{figure*}[t]
    \centering
    \includegraphics[width=\textwidth]{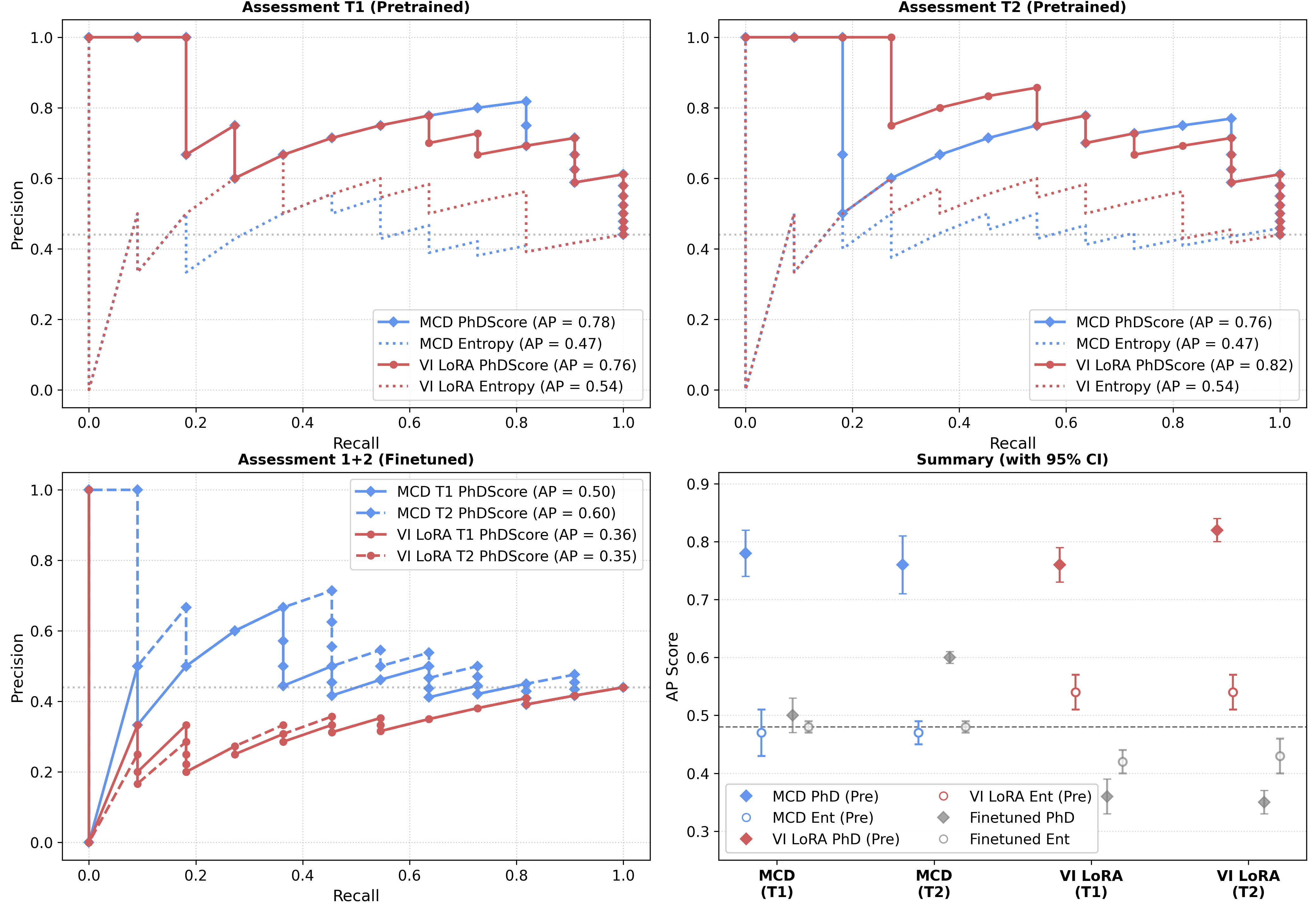}
    \caption{Longitudinal Clinical Validation. \textbf{Top Row:} Precision-Recall curves for Pre-trained models against Assessment 1 (Left) and Assessment 2 (Right). Solid lines (PhDScore) consistently outperform dotted lines (Entropy), with VI LoRA (Red) achieving the highest alignment (AP=0.82). \textbf{Bottom Left:} Fine-tuning collapses the correlation, indicating uncertainty resolution. \textbf{Bottom Right:} Summary of AP scores, highlighting the superiority of PhDScore over Entropy and the effect of fine-tuning.}
    \label{fig:clinical_validation_grid}
\end{figure*}

A crucial test of our method is whether the PhDScore captures genuine, clinically relevant articulatory challenges. Recent studies explored agreement between experts and models in transcription \cite{Kim2025ASRChildrenSSD}, but not phoneme-level speech pathology. For the BF-Sprache dataset, we performed a longitudinal validation using two formal logopedic reports (Assessment 1 and Assessment 2) conducted one year apart. This period covers substantial physiological change for the speaker. Figure \ref{fig:clinical_validation_grid} presents the Precision-Recall (PR) analysis of our uncertainty metrics against these expert assessments.

\noindent \textbf{PhDScore Outperforms Entropy.}
The Top Row of Figure \ref{fig:clinical_validation_grid} validates our core hypothesis: raw entropy is insufficient for identifying clinical difficulty. In both Assessment 1 and 2, the Entropy-based baselines (dotted lines) perform significantly worse than the PhDScore (solid lines). For VI LoRA on Assessment 2, the PhDScore achieves an Average Precision (AP) of 0.82, compared to just 0.54 for entropy. This confirms that incorporating historical error (PER) and stability (Agreement) into a composite score is essential for aligning model uncertainty with expert human perception.

\noindent \textbf{Robustness and Temporal Stability.} 
The strong correlation holds across both timepoints (Top Left vs. Top Right), demonstrating that the PhDScore captures persistent articulatory traits rather than transient noise. Notably, while VI LoRA (Red) and MCD (Blue) are very different methods of estimation uncertainty, both show excellent agreement with expert annotation. This suggests that modeling uncertainty in the parameter space or activation-based dropout capture a common signal which does also manifest in medical practice. Entropy alone on the other hand, is not capable of capturing this signal, indicated by significantly lower AP scores roughly at chance-level. This advantage of the PhDScore over Entropy is also implicitly confirmed in error rates by oversampling, shown in Table \ref{tab:ablation_sampling}.

\noindent \textbf{Resolution of Uncertainty.}
The Bottom Row illustrates the effect of personalization. After fine-tuning (Bottom Left), the strong correlation with clinical reports collapses (AP drops to $\approx 0.35$). The Summary Plot (Bottom Right) visualizes this dramatic shift: the high AP of the pre-trained model (colored markers) drops to near-random performance after fine-tuning (grey markers). This desired outcome confirms that the model has successfully learned the specific pathological patterns it was previously uncertain about, effectively "resolving" its epistemic uncertainty through our targeted oversampling strategy.

\noindent \textbf{Limitations.}
While this fine-grained, phoneme-level clinical validation is not feasible for public datasets like UA-Speech due to the lack of corresponding clinical reports, the consistent performance gains across 16 speakers, four intelligibility levels (Table \ref{tab:ua_speech_side_by_side_final}), and two typologically distinct languages  provide strong converging evidence that the uncertainty signal generalizes beyond a single speaker. The primary limitation remains the single-speaker nature of the BF-Sprache clinical validation, constrained by the difficulty of obtaining ethical approval for pediatric impaired speech. With recent ethical clearance, future work will expand this cohort to track developmental trajectories across multiple speakers and medical conditions.

\section{Conclusion}
\label{sec:conclusion}

We have presented a data-efficient personalization framework for ASR that uses a composite Phoneme Difficulty Score (PhDScore) to guide fine-tuning. By identifying a speaker's most challenging phonemes via Monte Carlo Dropout or VI LoRA and oversampling them during training, our method improves accuracy on non-normative speech. We also identified a clear trade-off between this deep personalization and generalization to normative speech, a key consideration for practical system design. Crucially, we demonstrated that our model-derived PhDScore, independently of the derivation method, aligns remarkably well with longitudinal clinical assessments from a speech therapist, validating its ability to robustly identify core articulatory challenges over time. The subsequent disappearance of this correlation after fine-tuning confirms that our method effectively resolves the model's uncertainty through targeted learning. This work represents a practical step toward creating more effective, interpretable, and truly personalized ASR systems, with potential applications in both assistive technology and as a supplemental tool for clinical practice.

% \clearpage

% \section{Acknowledgments}

\section{Generative AI Use Disclosure}
Generative AI tools were utilized during the preparation of this manuscript for linguistic refinement of the text and to optimize the Python scripts used for data visualization. Furthermore, Large Language Models were employed as a data-processing aid to identify and validate semantically meaningful sentence patterns required for the semantic re-chaining process.

\footnotesize
\bibliographystyle{IEEEtran}
\bibliography{mybib}

\end{document}